\documentclass[twocolumn,showpacs,amsmath,amstex,amssymb, mathfonts,floats]{revtex4}
\pdfoutput=1
\usepackage{amsthm,color,amsfonts,graphicx,verbatim,psfrag}
\usepackage{amsmath}
\usepackage{amssymb}
\usepackage{amsthm}
\usepackage{amsfonts}
\usepackage{listings}
\usepackage{enumerate}
\usepackage{latexsym}
\usepackage{psfrag}
\usepackage{bm}
\usepackage{graphicx}
\usepackage{dsfont}

\usepackage{hyperref}
\hypersetup{
    bookmarks=true,         
    unicode=false,          
    pdftoolbar=true,        
    pdfmenubar=true,        
    pdffitwindow=false,     
    pdfstartview={FitH},    
    pdftitle={Many-body localization in disorder-free systems},    
    pdfauthor={},     
    pdfsubject={},   
    pdfcreator={},   
    pdfproducer={}, 
    pdfkeywords={keyword1} {key2} {key3}, 
    pdfnewwindow=true,      
    colorlinks=true,       
    linkcolor=black,          
    citecolor=blue,        
    filecolor=magenta,      
    urlcolor=blue           
}

\newcommand{\be}{\begin{equation}}
\newcommand{\ee}{\end{equation}}
\newcommand{\bea}{\begin{eqnarray}}
\newcommand{\eea}{\end{eqnarray}}

\renewcommand{\phi}{\varphi}
\renewcommand{\epsilon}{\varepsilon}

\begin{document}


\title{Many-body localization in disorder-free systems: the importance of finite-size constraints}

\author{Z. Papi\'c}
\address{School of Physics and Astronomy, University of Leeds, Leeds, LS2 9JT, United Kingdom}
\address{Perimeter Institute for Theoretical Physics, Waterloo, ON N2L 2Y5, Canada}

\author{E. Miles Stoudenmire}
\address{Perimeter Institute for Theoretical Physics, Waterloo, ON N2L 2Y5, Canada}

\author{Dmitry A. Abanin}
\address{Department of Theoretical Physics, University of Geneva, 24 quai Ernest-Ansermet, 1211 Geneva, Switzerland}
\address{Perimeter Institute for Theoretical Physics, Waterloo, ON N2L 2Y5, Canada}

\date{\today}
\begin{abstract}
Recently it has been suggested that many-body localization (MBL) can occur in translation-invariant systems, and candidate 1D models have been proposed. We find that such models, in contrast to MBL systems with quenched disorder, typically exhibit much more severe finite-size effects due to the presence of two or more vastly different energy scales. In a finite system, this can easily create an artificial splitting of the density of states (DOS) into bands separated by large energy gaps. We argue that in order for such models to faithfully represent the physics of the thermodynamic limit, the ratio of the relevant coupling parameters must be larger than a certain cutoff that depends on system size, and should be chosen in such a way that various bands in the DOS of a given model overlap with one another. By setting the parameters in this way to minimize the finite-size effects, we then perform exact diagonalization studies of several translation-invariant MBL candidate models. Based on the variety of diagnostics, including entanglement properties and the behaviour of local observables, we find the systems exhibit thermal (ergodic), rather than MBL-like behaviour. Our results suggest that MBL in translation-invariant systems with two or more very different energy scales is less robust than perturbative arguments suggest, possibly pointing to the importance of non-perturbative effects which induce delocalization in the thermodynamic limit. 

\end{abstract}


\maketitle

\section{Introduction}

\begin{figure}[hhh]
\begin{center}
\includegraphics[width=\columnwidth]{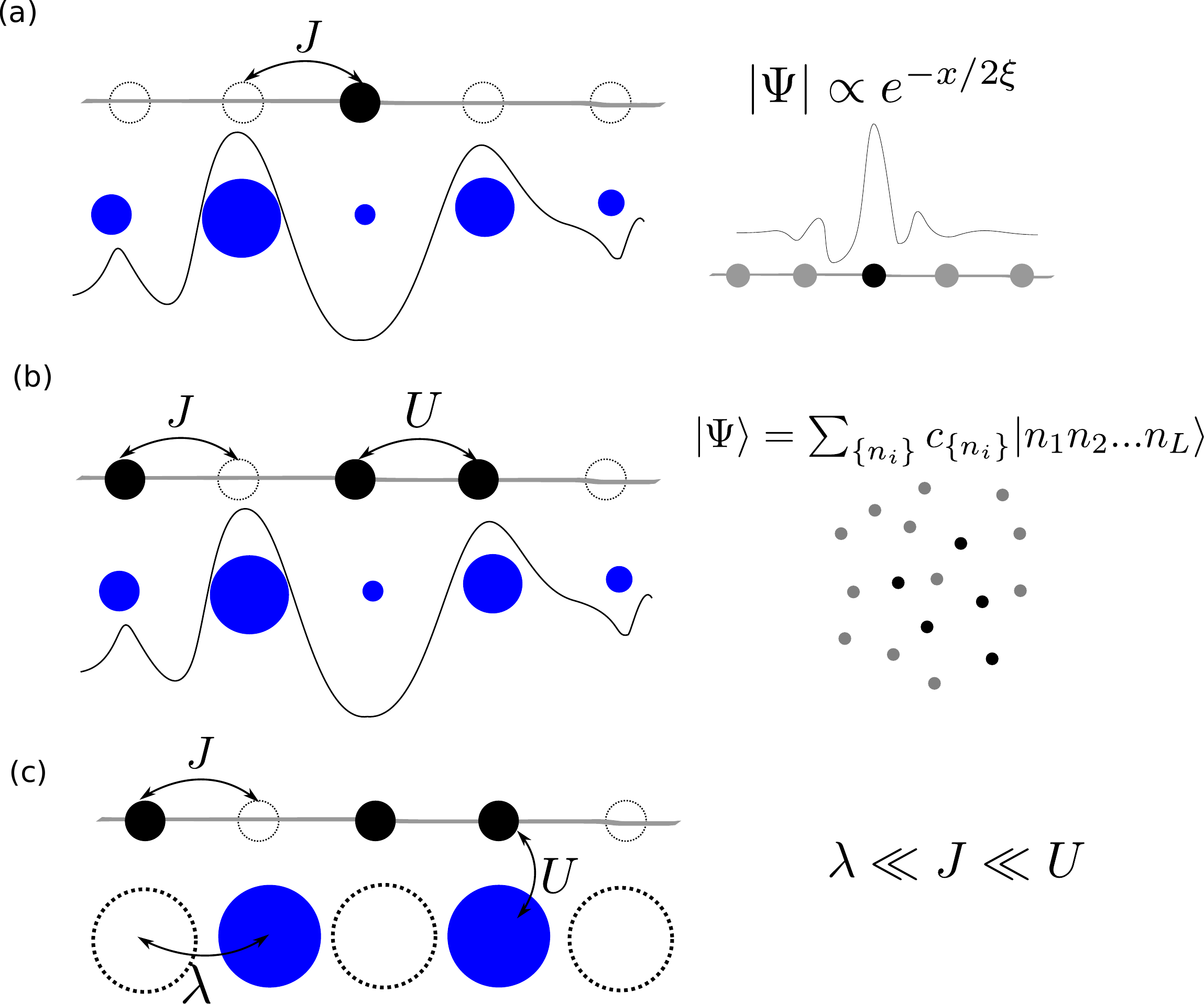}
\caption{ (Color online) (a) Anderson localization of a single particle hopping on a lattice in the presence of external disorder potential (blue circles). All eigenstates of such a particle in 1D are exponentially localized in space. (b) Many-body localization of particles hopping on a 1D lattice and interacting between nearest neighbour sites. In this case also, the particles move in the background of static external disorder potential (blue circles). For sufficiently strong disorder, the wave function of typical eigenstates is localized (or weakly entangled) in the many-body Fock space. (c) A proposal for MBL in a translation-invariant system, similar to the 1D Hubbard model. In this case, localization is expected to arise dynamically from strong on-site interaction between two species of particles with very different masses ($\lambda\ll J \ll U$), although no external randomness is imposed on the system. \label{FigIntro}}
\end{center}
\end{figure}

One of the remarkable consequences of quantum mechanics is the localization of a single particle moving in a disordered medium in one and two spatial dimensions~\cite{Anderson58}. This phenomenon, known as Anderson localization, relies on the presence of a static ``quenched" disorder landscape through which a particle can hop, causing its eigenstates to form localized wave packets instead of extended (Bloch) states in a usual crystal [see Fig.~\ref{FigIntro}(a)]. This has dramatic consequences for dynamics and transport~\cite{MacKinnon}.

Recently it was found that localization can persist in interacting systems of many particles~\cite{Basko06, Mirlin05, Oganesyan07, Pal10}. One of the simple examples of such systems is shown in Fig.~\ref{FigIntro}(b); it consists of a 1D lattice half-filled with particles that can hop as well as interact between nearest neighbour sites. Moreover, similar to the Anderson case, the system is in the background of an external disorder potential that is considered static and only enters the description as a random chemical potential on each site. Recent work~\cite{Basko06, Mirlin05, Oganesyan07, Pal10} has shown that for sufficiently strong disorder such a system enters a ``many-body localized" (MBL) phase  where, similar to the Anderson case, transport is inhibited even at non-zero temperature. While in the Anderson case the localization directly implies spatial locality of the wave function of typical eigenstates, in the interacting case each eigenstate $|\Psi\rangle$ is a linear combination of Fock states $|n_1,n_2,...,n_L\rangle$, where $n_i$ is the number operator on site $i$. There are exponentially many possible $|\{n_i\}\rangle$ and in a thermal (delocalized) system, $|\Psi\rangle$ has a non-zero component on almost all of them. When disorder is strong enough, it turns out that a typical $|\Psi\rangle$ has non-zero projection on a far fewer $|\{n_i\}\rangle$. This is a signature of many-body localization, sometimes referred to as ``localization in Fock space" because of the system's inability to explore the entire many-body Hilbert space.       

A special structure of eigenstates $|\Psi\rangle$ in the MBL phase generally is a result of an emergence of an extensive number of locally conserved quantities~\cite{Serbyn13-2, Huse13, Imbrie14, Ros14,  Chandran14}. The appearance of these local integrals of motion is responsible for the ergodicity breaking in the MBL phase, and leads to a characteristic entanglement structure~\cite{Vosk13, Serbyn13-2, bauer} of its individual many-body eigenstates. In ergodic (thermalizing) systems, the entanglement of typical eigenstates at high energy densities above the ground state obeys the so-called ``volume law": the von Neumann entropy of a large finite subsystem in such an eigenstate scales as the total number of degrees of freedom in the subsystem. This is simply a manifestation of $|\Psi\rangle$ being a random linear combination of nearly all Fock states $|\{n_i\}\rangle$. In contrast, the extensive number of local integrals of motion (which span the entire many-body Hilbert space) constrains the eigenstates in the MBL phase to have significantly ``less" entanglement: their von Neumann entropy only scales as the \emph{area} of the subsystem (therefore, in 1D the entanglement entropy is a constant). Further, dynamical probes such as the spreading of entanglement entropy following a global quench~\cite{Prosen08, Moore12,Serbyn13-1}, as well as the time evolution of local observables~\cite{deer, Vasseur14, Serbyn14}, show that MBL phases have universal properties that distinguish them from both the ergodic phase and the non-interacting Anderson insulator. 

One of the outstanding questions is whether the MBL phase necessarily requires disorder, i.e. could MBL-like physics and ergodicity breaking arise in translation-invariant systems, solely as a consequence of interactions? Recent works suggest that this intriguing phenomenon may indeed be possible~\cite{kagan1, kagan2, carleo, muller1, muller2, deroeck1, deroeck2, deroeck3, garrahan, grover, garrison, yao}. 
A generic class of proposed translation-invariant MBL models involves two species of fermions $a$ and $b$ hopping on a 1D lattice
\begin{eqnarray}\label{model1}
\nonumber H &=& -J\sum_i a_i^\dagger a_{i+1} -\lambda\sum_i b_i^\dagger b_{i+1} + h.c. \\
&+& \sum_{k,l} U(k-l) a_k^\dagger a_{k+\sigma} b_l^\dagger b_{l+\sigma'}.
\end{eqnarray}
Here $J$ represents the hopping amplitude of fast ($a$) particles and $\lambda \ll J$ is the hopping of heavy ($b$) particles. The particles also interact via $U(k-l)$. In the simplest case, $\sigma=\sigma'=0,k=l$ and the interaction reduces to an on-site density-density term, thus the model is formally identical to the 1D Hubbard model~\cite{yao}. This model is schematically shown in Fig.~\ref{FigIntro}(c) and can be viewed as a generalization of MBL models when the disorder is no longer static. That is, the particles that generate the disorder are included in the system and allowed to undergo their own quantum dynamics and interact with the original particles of the system. Apart from the 1D Hubbard model, we will also consider an alternative model with $\sigma=1,\sigma'=0,k=l$ which can be visualized as light particles hopping on a lattice subject to the kinematic constraint depending on whether the hop extends across a heavy particle (``barrier") or not~\cite{muller1}. Finally, we will also study a related one-component model which is the limiting case of the model $\sigma=1,\sigma'=0$, where light particles have been integrated out, leading to an effective single species model with long-range interaction~\cite{muller2}. 

The expectation is that heavy particles in such models dynamically generate an effective disorder potential that localizes light particles. The possibility of such systems to display true~\cite{muller1, muller2} or ``partial"~\cite{yao} MBL has recently been discussed. It was also argued that, regardless of whether such systems are truly MBL or not, they may break ergodicity in subtle ways that can only be detected by sophisticated probes such as the ``post-measurement" entanglement of their eigenstates~\cite{grover, garrison}.

In this paper we critically examine the mentioned family of translation-invariant candidate models for MBL. In Section~\ref{sec:2comp} we consider models with two species of fermions (the 1D Hubbard model and the model with particles and barriers), while Section~\ref{sec:1comp} focuses on a related single-particle model. Because the general theoretical understanding of such models is currently lacking, our approach combines insights from analytic solutions of small systems with exact diagonalization of larger finite systems. Our conclusions are presented in Section~\ref{sec:conc} and can be briefly summarized as follows.

The main observation is that, in contrast to MBL models with quenched disorder, the translation-invariant models possess two or more drastically different energy scales (e.g., the hopping amplitudes of light and heavy particles). We argue that this leads to pronounced finite-size effects, and implicitly places constraints on the values of parameters of the models (e.g., the minimal mass ratio of light and heavy particles). For example, for a given Hubbard chain of size $L$ and fixed $J, U$, there exists a cutoff $\lambda_c(L)$ such that only values of $\lambda>\lambda_c(L)$ yield a faithful representation of the system in the thermodynamic limit. For values $\lambda \ll \lambda_c(L)$, the system's DOS artifically splits into discrete bands separated by large energy gaps. The suppressed overlap between bands in DOS means that the effective Hilbert space is drastically reduced. In this regime, a finite system indeed displays MBL-like behaviour, as pointed out previously~\cite{yao}. We argue that this feature, however, is a finite-size effect that may not persist in the thermodynamic limit, as the broadening of each band is extensive in system size. Thus, the parameters of the model (e.g., $\lambda$) have to be chosen carefully for a given finite system to avoid the mentioned DOS splitting. For such choices of $\lambda$, we find that the typical behaviour of the system -- diagnosed by the entanglement measures and thermalization of local observables~\cite{deutsch, srednicki, Rigol08} -- appears to be ergodic rather than localized. Our results show that to demonstrate MBL in translation-invariant systems one needs to carefully examine the finite sample's DOS for the given hopping and interaction energy scales. More generally, the results imply that translation-invariant MBL is significantly more fragile than the disorder-driven one, and may generically exhibit a crossover to the thermal behaviour in the thermodynamic limit. 

\begin{figure}[htb]
\begin{center}
\includegraphics[width=\columnwidth]{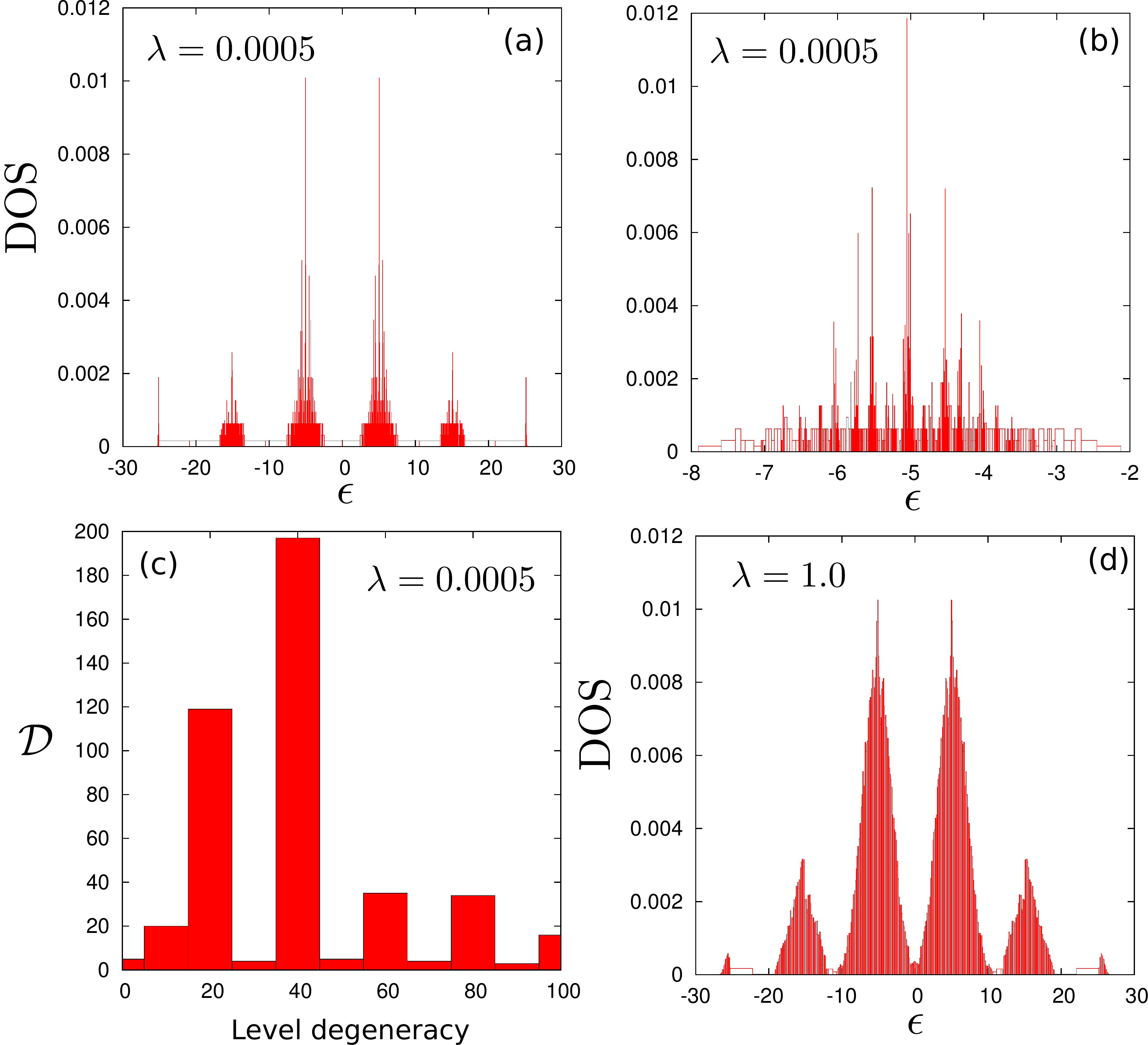}
\caption{ (Color online) The DOS for the model (\ref{eq:yao}) and separation into non-overlapping bands for $L=10$, $U=10$, $J=1$.\label{Fig1}}
\end{center}
\end{figure}

\section{Models with two species of particles}\label{sec:2comp}

We first consider a 1D Hubbard model with two species of fermions $a, b$ interacting on the same site, which is equivalent to the spin ladder model studied in Refs.~\cite{Paredes05,yao}:
\begin{eqnarray}\label{eq:yao}
H=\sum_i -Ja_i^\dagger a_{i+1} - \lambda b_i^\dagger b_{i+1} + h.c. + U \rho_i^a \rho_i^b,
\end{eqnarray}
where $\rho_i^\sigma=\sigma_i^\dagger\sigma_i-\frac{1}{2}$, $\sigma=a,b$. The model is depicted in Fig.~\ref{FigIntro}(c) and conserves the total number of particles of each species. We consider periodic boundary condition ($i+L\equiv i$) and half filling for both $a$ and $b$ particles. We are interested in the regime $\lambda\ll J\ll U$ where the system was conjectured to exhibit MBL behaviour~\cite{yao}. We first discuss how the DOS in a finite system differs from thermodynamic limit, and from this derive the conditions on the parameters of the model that minize the finite-size effects [Section~\ref{sec:dos}]. In Section~\ref{sec:ent} we relate the behaviour of entanglement and particle fluctuations to the discussed DOS and estimates derived in Section~\ref{sec:dos}.

\subsection{|Finite system's DOS and constraints on the parameters of the model}\label{sec:dos}

The eigenstates of the model (\ref{eq:yao}) have a simple structure when particles are immobile ($J=\lambda=0$). In this case, $a$ and $b$ particles are randomly placed on lattice sites, and the energy is determined solely by $U$. The DOS in this case consists of massively degenerate bands that are separated by energy $U$. Next, let us turn on $0<J\ll U$ while keeping $\lambda=0$. Now, $a$ particles are Anderson-localized in the effective disorder profile generated by the random positions of $b$ particles. As a result, a band with a given classical interaction energy splits into many ``mini-bands". A typical mini-band still has exact degeneracy of at least $2L$, owing to the translation and reflection symmetry, and the typical energy difference between neighbouring mini-bands is of the order $\Delta E(L)$, which decays exponentially in the thermodynamic limit. For $J/U\lesssim 0.1$ and the system sizes available numerically  ($L\leq 10$), the bands with different classical energies remain well-separated. In principle, this is a finite-size effect that places the system far from the thermodynamic limit (effectively, because the system is too small to absorb an energy of the order $U$).

To illustrate the above, in Fig.~\ref{Fig1} we show an example of the DOS for $L=10$ site system and $J/U = 0.1$. Panel (a) is for the case $\lambda=0.0005$ where classical bands can be clearly identified. The bands are quasidegenerate because of non-zero value of $\lambda$; for $\lambda=0$ the bands indeed shrink to delta function peaks. In panel (b) we zoom in on one of the classical bands centered around $\epsilon=-5$. As stated above, we can now identify its fine structure corresponding to numerous mini-bands. Each of the mini-bands is at least $2L$ degenerate, as we can see in panel (c) where we count the degeneracy of energy levels and plot the distribution of degeneracies (the levels are taken to be degenerate if their energy difference is less than $10^{-4}$). We also see that our estimate of typical degeneracy being $2L$ is likely a conservative one because the distribution is peaked around the larger value which in general is some multiple of $2L$ [e.g., $4L$ in Fig.~\ref{Fig1}(c)]. For comparison, in Fig.~\ref{Fig1}(d) we show the DOS for $\lambda=1.0$ where the classical energy bands start to overlap.    

\begin{figure}[htb]
\centering
\includegraphics[width=0.4\textwidth]{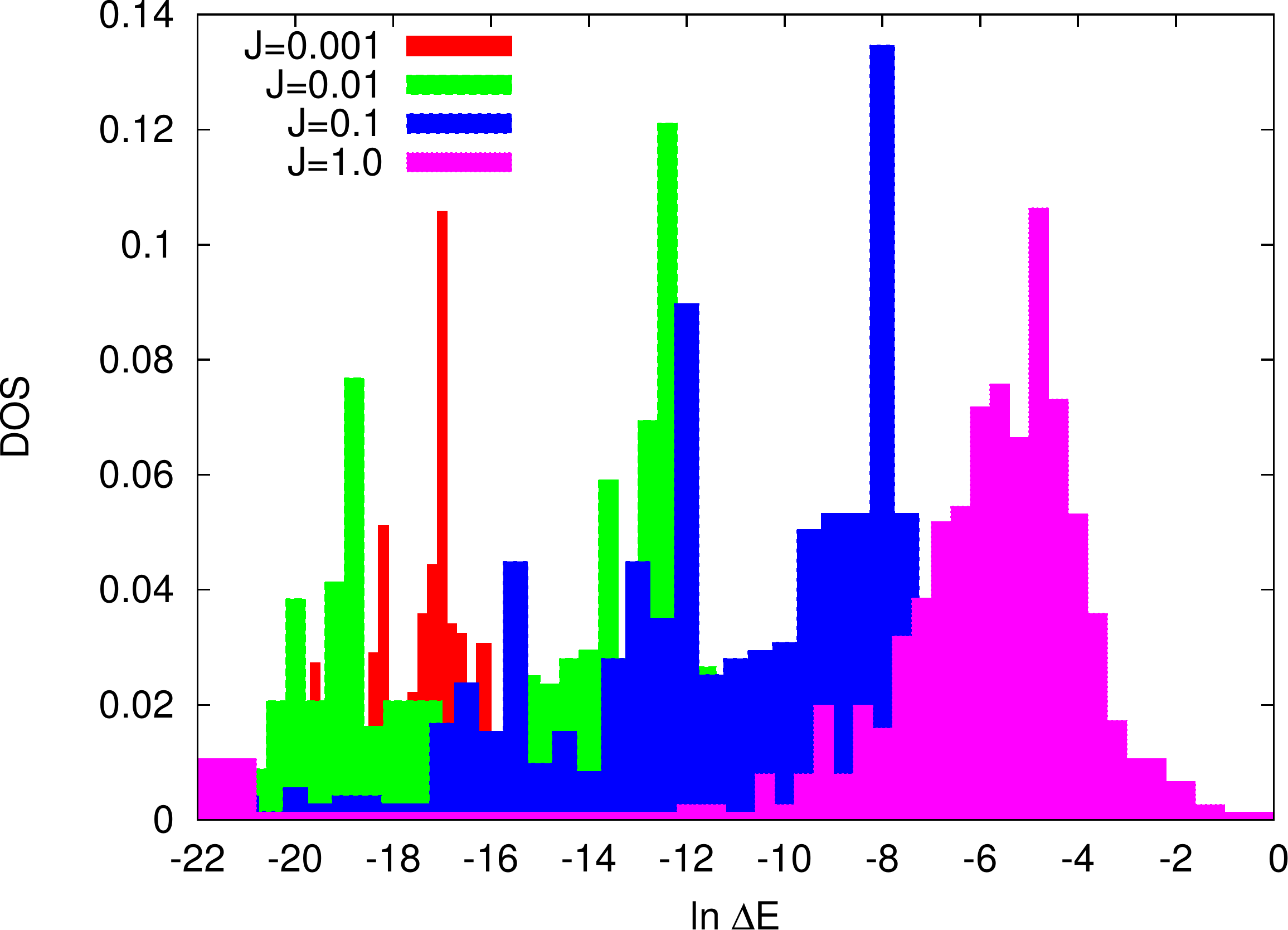}
\caption{(Color online) The distribution of level spacings inside the central band of $L=10$ system. We set $U=10$ and $\lambda=0$, and assume periodic boundary conditions. The locations of the peaks are roughly consistent with the analytic estimate in Eq.~(\ref{eq:estimate}).}
\label{FigA0}
\end{figure}

Let us provide an estimate for the critical $\lambda$ at which the DOS bands start to overlap. As each band contains an exponentially large number of states, turning on finite $\lambda$ can lead to thermalization within each band. We note that, since the bands are separated by an energy of order $U$, the hopping $\lambda$ typically couples states within the same band with an amplitude $\lambda J/U$ (since for many states moving a heavy particle changes the classical energy by $U$). If $\lambda$ is chosen very small, such that $\lambda J/U$ is smaller than the mini-band spacing $\Delta E(L)$, then the effect of turning on the hopping is perturbative, and it cannot significantly modify the eigenstates. For instance, in our example in Fig.~\ref{Fig1}(a) we find the entanglement entropy of the eigenstates still obeys the area law. This is consistent with the system being in an MBL phase, but it is also likely that the finite system is then not reproducing the behaviour of the thermodynamic limit. Thus, to obtain a faithful representation of the thermodynamic limit in a large finite system of size $L$, we require $\lambda$ to be sufficiently large to hybridize several mini-bands
\begin{equation}
\lambda J/U\gtrsim \Delta E(L).
\end{equation} 
To estimate $\Delta E(L)$, consider the largest band with $n(L)={L \choose L/2} {L/2 \choose \left \lfloor{L/4}\right \rfloor }^2$ states. At $\lambda=0$, a conservative estimate for the broadening of this band is $(J^2 /U)\sqrt{L}$. We can easily convince ourselves that the broadening for $\lambda=0$ is given by $\sim J^2/U$ by exactly solving a small system, e.g. $L=4$. Because of translation symmetry, we can restrict to a sector with the given momentum (e.g., $k=0$). In this sector, there are 10 states whose exact eigenenergies are
\begin{eqnarray}
0, \; \pm 2J, \; \pm \sqrt{U^2+4J^2}, \; \frac{1}{2}(\pm U \pm \sqrt{U^2 + 16 J^2}), 
\end{eqnarray}
thus the extremal eigenvalues are indeed shifted by $\sim L(J^2/U)$. In a large system, assuming normal distribution for the modified energy values once we turn on $J$ (while still keeping $\lambda=0$), we obtain the rough estimate for the broadening 
\begin{equation}\label{eq:estimate}
\Delta E(L)\sim \frac{2J^2L^{3/2}}{U n(L)},
\end{equation} 
where we took into account the degeneracy of $2L$ of a typical mini-band. For $L=10$ this gives $\Delta E(L)\sim  2.5\times 10^{-2} J^2/U$, which is of correct order but, as expected, underestimates the exact result. Thus, the necessary (but possibly not sufficient) condition for the system to be able to capture the thermodynamic limit is 
\begin{equation}
\lambda \gtrsim \lambda_c\approx U \Delta E(L)/J \approx 0.025 J.
\end{equation} 
In the following, we fix $U=10, J=1$. Signatures of MBL were found~\cite{yao} for such parameters, when $\lambda$ was taken very small ($10^{-3}-10^{-2}$). Our estimate above gives $\lambda_c\approx 0.025$; choosing $\lambda$ as small as $10^{-3}$ would necessitate very large system sizes to make sure that finite-size effects are eliminated. 

\begin{figure}[htb]
\begin{center}
\includegraphics[width=0.9\columnwidth]{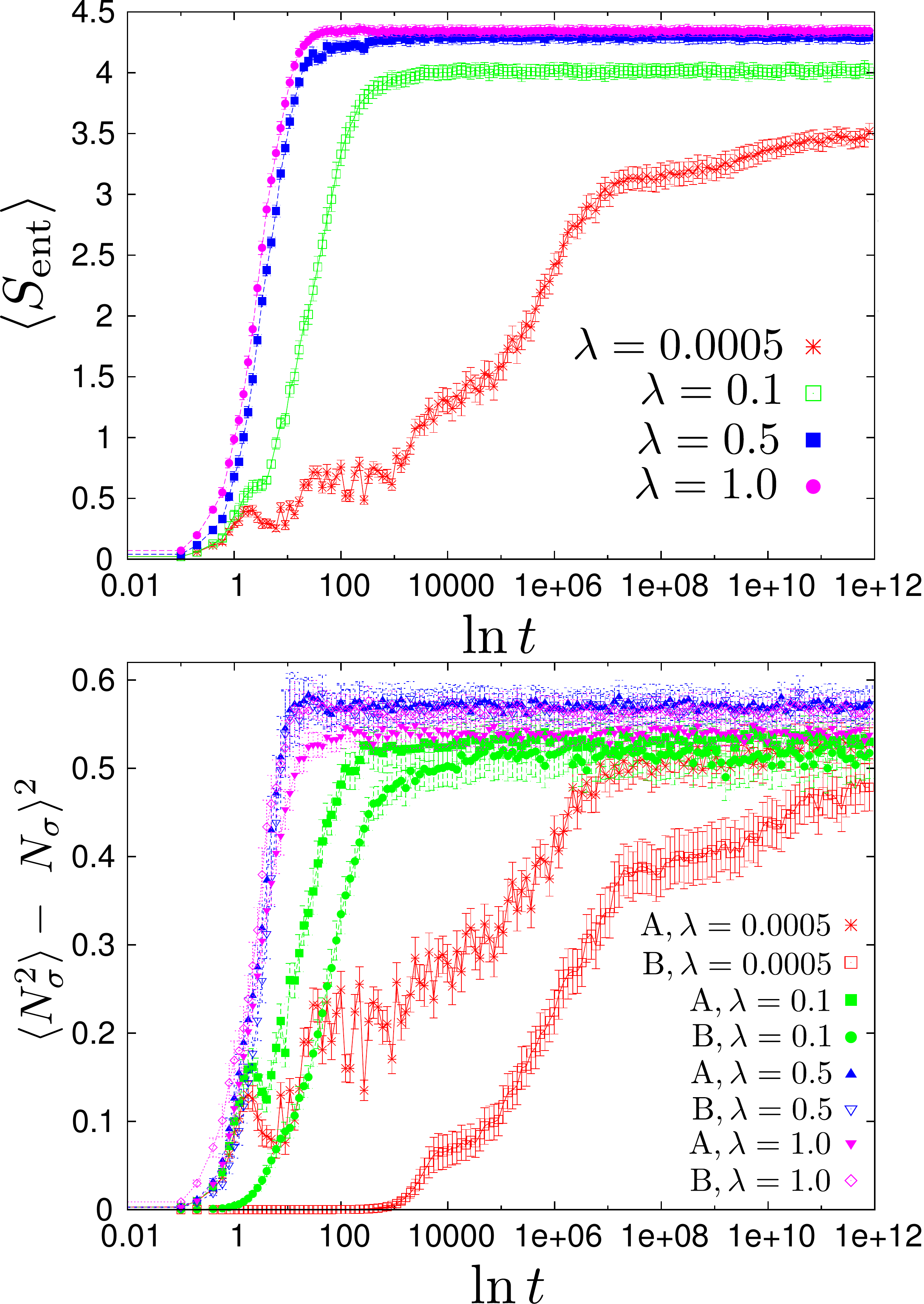}
\caption{ \label{Fig2} (Color online) Growth of entanglement entropy (left) and particle number fluctuations (right) in one half of the system as a function of time. System is $L=8$ sites at half filling of species $a$ and $b$.}
\end{center}
\end{figure}

In Fig.~\ref{FigA0} we test the derived estimate in Eq.~(\ref{eq:estimate}) with the result obtained from exact diagonalization. We plot the distribution of the level spacings between the minibands within the central (largest) band of the $L=10$ system (we set $U=10$ and $\lambda=0$). This central band contains $n(10)=2520$ states whose energy differences are plotted in Fig.~\ref{FigA0}. Level spacings between symmetry-related states have been removed, and the remaining ones are plotted on a log scale for easier comparison with our analytic estimate of $\Delta E(L)$ in Eq.~(\ref{eq:estimate}). For the given choices of $J$, Eq.~(\ref{eq:estimate}) predicts $\ln \Delta E_{J=0.001} \approx -19.8$, $\ln \Delta E_{J=0.01} \approx -15.2$, $\ln \Delta E_{J=0.1} \approx -10.6$ and $\ln \Delta E_{J=1.0} \approx -6.0$. These values are consistent with, though somewhat lower than the exact peaks of the distribution in Fig.~\ref{FigA0}.

\subsection{Entanglement and particle fluctuations as a diagnostics of MBL}\label{sec:ent}

Having derived the rough estimates for the parameter range in the Hubbard model, we proceed to compute the well-known diagnostics of many-body localization. One particularly useful probe is performing a global quench and examining the spreading of correlations after the system is initialized in a product state at time $t=0$ and unitarily evolved with the Hamiltonian (\ref{eq:yao}) for $t>0$. We will discuss the dynamical behaviour of the system undergoing a quench for the parameter values obtained in Section~\ref{sec:dos}.

In Fig.~\ref{Fig2} we compute the evolution of entanglement entropy $S_{\rm ent}$ under the global quench, as well as the particle number fluctuations $\langle N_{\sigma}^2\rangle - \langle N_{\sigma} \rangle^2$, $\sigma=a,b$, in one half of the system. Data is averaged over random initial product states. 
For an MBL system, we expect a slow, logarithmic in time growth of $S_{\rm ent}$, and a much slower growth of particle fluctuations~\cite{Moore12}. This behaviour reflects the slow dephasing and the suppression of transport, respectively. For the smallest value of $\lambda$ (roughly corresponding to Ref.~\cite{yao}) we indeed find a signal that is reminiscent of MBL physics in the quenched disorder case~\cite{Prosen08, Moore12, Serbyn13-1}. The particle fluctuations, especially at long times, are significantly larger, which suggests the system becomes diffusive at very long times. However, for $\lambda \gtrsim 0.1$, where according to our estimate in Eq.~(\ref{eq:estimate}) the system should be ``closer" to the thermodynamic limit, we find signatures of thermal behaviour. For example, the growth of entanglement is significantly faster and essentially featureless, with no indication of long time scales that characterize the transient regime. The saturation value of the entropy is high and approaches the thermal value independent of $\lambda$ for large $\lambda=0.5, 1$, while particle number fluctuations grow at a similar rate for both species. These results suggest that as long as the DOS consists of overlapping mini-bands, the generic behaviour of the system is diffusive, and signatures of MBL only appear at intermediate time scales.   

We have reached similar conclusions by studying the model with particles and barriers proposed in Ref.~\cite{muller1}:
\begin{eqnarray}\label{eq:mullertwocomponent}
H=\sum_i -J (a_i^\dagger a_{i+1} + h.c.)(1-b_i^\dagger b_i) - \lambda b_i^\dagger b_{i+1} + h.c.
\end{eqnarray}
This is the special case $\sigma=1, \sigma'=0$ of Eq.~(\ref{model1}). In this model there is a kinematic constraint on the hopping of species $a$ depending on whether the hop extends across a $b$ particle or not. We have studied this model and found it displays similar phenomenology to the model in Eq.~(\ref{eq:yao}). These two-component models [Eqs.~(\ref{eq:yao}) and (\ref{eq:mullertwocomponent})], however, have the disadvantage of a rapid increase of Hilbert space dimension with system size $L$. Therefore, in the following section  we focus on a single-component model~\cite{muller2} that is expected to capture similar physics, while at the same time allowing the numerics to reach bigger system sizes.  

\begin{figure}[htb]
\begin{center}
\includegraphics[width=\columnwidth]{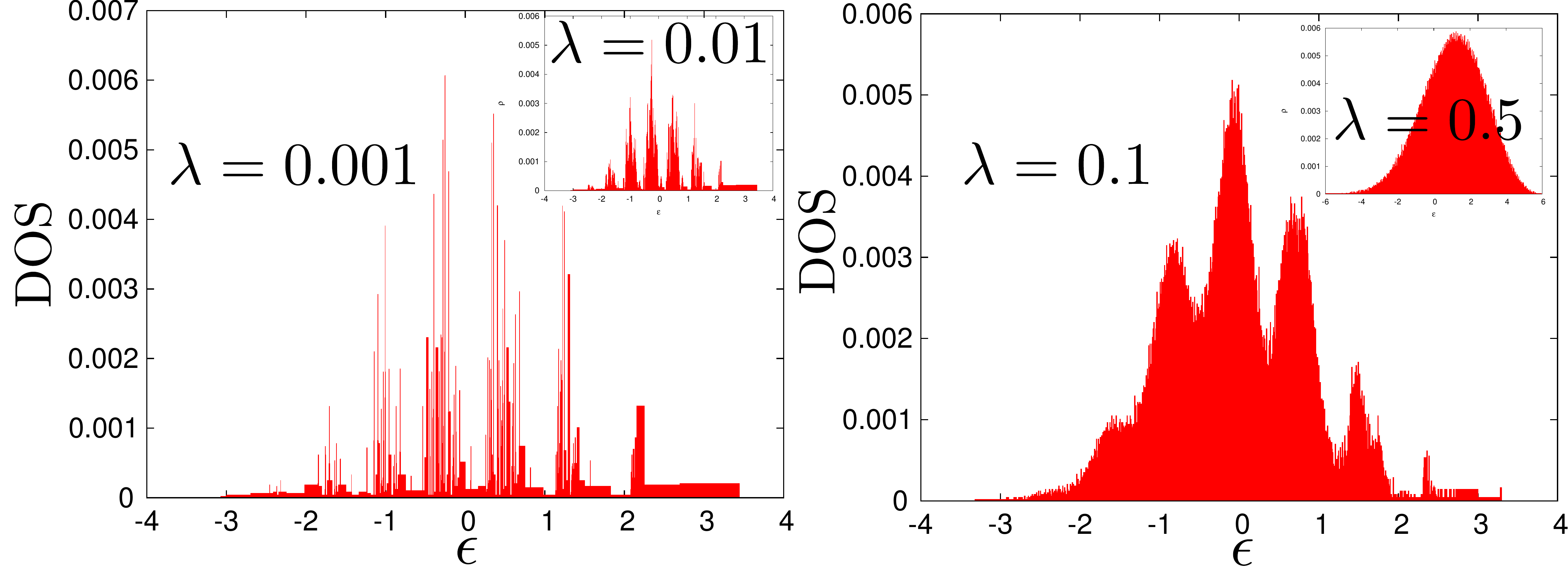}
\caption{ (Color online) The DOS for an $L=18$ site chain at half filling described by the model (\ref{eq:muller}). Decoupled bands are visible for small $\lambda=0.001,0.01$ (left), they begin to mix around $\lambda_c=0.1$ and fully merge by $\lambda=0.5$ (right). \label{Fig4} }
\end{center}
\end{figure}

\section{A single-species model}\label{sec:1comp}

Starting from the model (\ref{eq:mullertwocomponent}), the strategy is to integrate out light particles $a$ and obtain an effective single-component model for $b$ particles only. This is possible under the approximation that there is exactly one light particle between the heavy ones~\cite{muller2}. In this case, we arrive at the following Hamiltonian:
\begin{eqnarray}\label{eq:muller}
H = -\lambda \sum_i b_i^\dagger b_{i+1} + h.c. + n_i \sum_{r>0} \frac{U}{r^\beta} n_{i+r} \prod_{k=1}^{r-1} (1-n_{i+k}),  
\end{eqnarray} 
where $n_i = b_i^\dagger b_i$. Below we set $U=1$ and $\beta=2$~\cite{muller2}.

At $\lambda=0$, the DOS of the model (\ref{eq:muller}) consists of several bands separated by gaps of order one. Each band contains configurations with the same number of occupied pairs of nearest-neighbour sites. The interaction energy is not exactly the same, as it depends on how the occupied pairs of sites are arranged; thus, bands further split into mini-bands. In contrast to the Hubbard model, in this case the mini-bands can have higher degeneracy. Fig.~\ref{Fig4} shows the DOS for $L=18$ site chain at half filling and different values of $\lambda$.  For small $\lambda=0.001, 0.01$ (Fig.~\ref{Fig4},left), the DOS consists of non-overlapping bands. By $\lambda=0.1$ these band start to mix, and for $\lambda=0.5$ the DOS becomes contiguous. We therefore expect that critical $\lambda_c$ in this case lies in the interval (0.01;0.1). In the following we focus on three cases: $\lambda=0.001$, 0.1 and 0.5, and test whether local observables satisfy the ``eigenstate thermalization hypothesis"~\cite{deutsch, srednicki, Rigol08, Kim_ETH} (ETH), and study the scaling of entanglement in the eigenstates and its growth following a quench from an initial product state.

\begin{figure}[htb]
\centering
\includegraphics[width=0.9\linewidth]{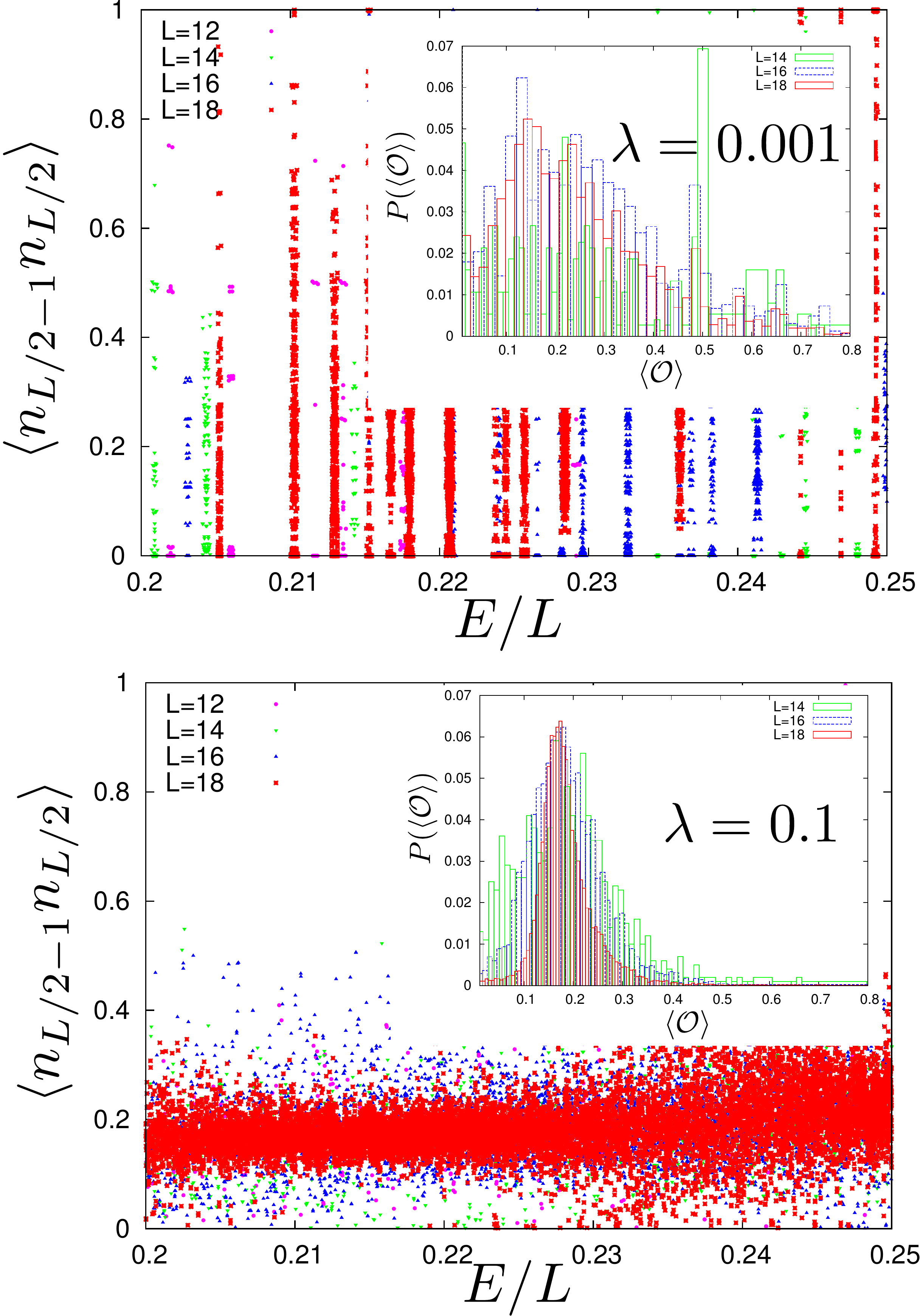}
\caption{(Color online) Testing the ETH for the model (\ref{eq:muller}) at half filling. Upper panels: Expectation value of the local operator $\mathcal{O}\equiv n_{L/2-1}n_{L/2}$ in many-body eigenstates in the energy window  $0.2<E/L<0.25$, for several values of $\lambda$. Bottom panels: the distribution of $\langle \cal O\rangle$ for the same values of $\lambda$.}
\label{FigA1}
\end{figure}
In Fig.~\ref{FigA1} we test the ETH for the local operator $\mathcal{O}\equiv n_{L/2-1}n_{L/2}$ acting on the middle two sites in the chain. We compute the expectation value $\langle \cal O\rangle$ for the many-body eigenstates in the region $0.2<E/L<0.25$ near the middle of the band, for $\lambda=0.001$ and $\lambda=0.1$. Different choices of local operators $\cal O$ are possible, but yield qualitatively the same results. In the insets of Fig.~\ref{FigA1} we also show the distribution of $\langle \cal O\rangle$. For $\lambda=0.001$ the system is strongly non-ergodic, with a very broad distribution of $\langle \cal O \rangle$ [Fig.~\ref{FigA1}, top]. At $\lambda=0.1$ the system behaves according to the ETH prediction: the distribution of local observables becomes increasingly sharper with system size, and most of the eigenstates behave typically. 

\begin{figure}[htb]
\begin{center}
\includegraphics[width=0.9\columnwidth]{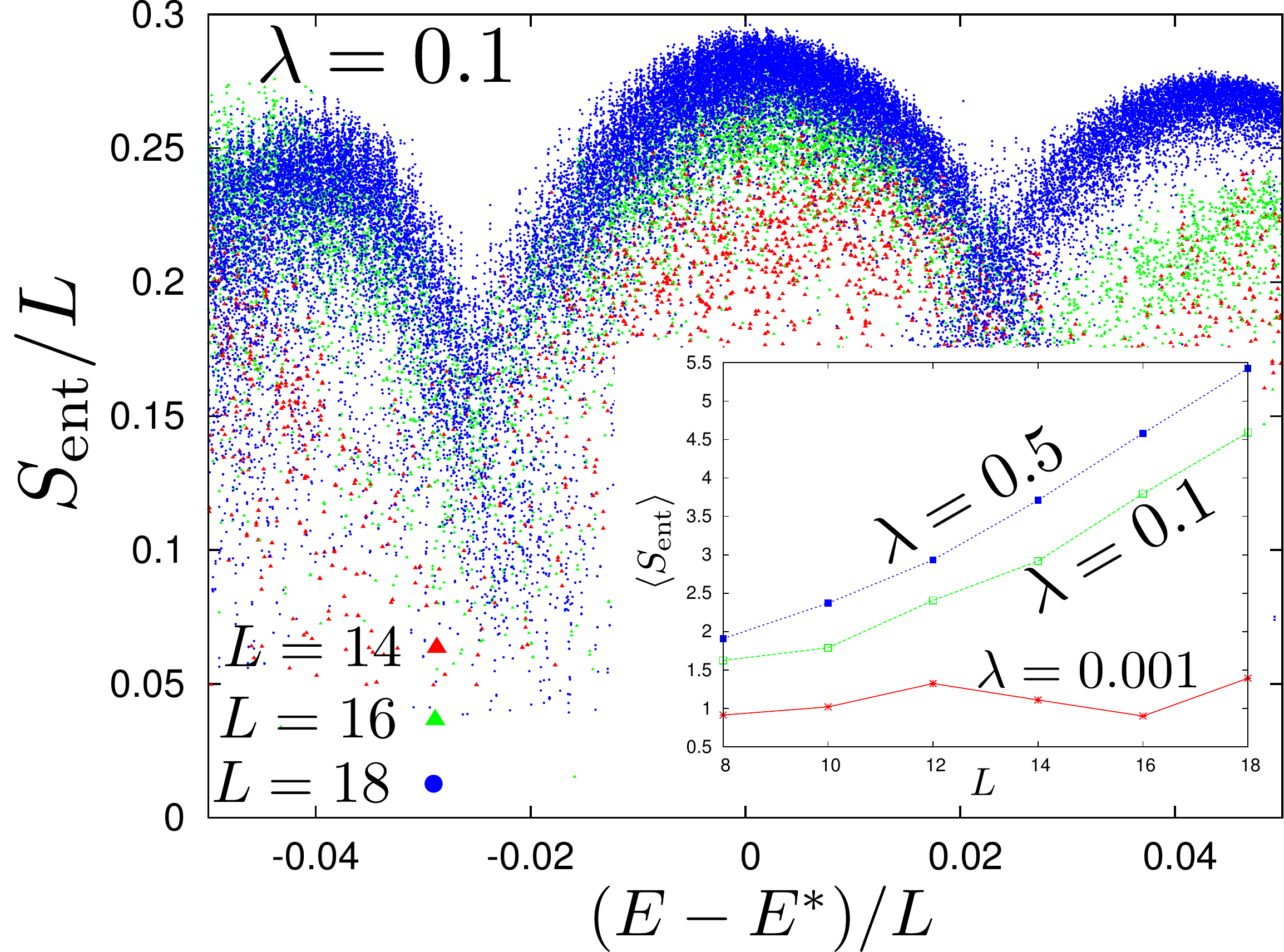}
\caption{ (Color online) Entanglement entropy density of the eigenstates in the middle of the band for various system sizes and $\lambda=0.1$ (left). Data for different $L$ is shifted by $E^*$ so that zero energy corresponds to the center of the band where states have the largest $S_{\rm ent}$. Average $S_{\rm ent}$ in the window $|E-E^*|/L<0.02$ scales extensively for $\lambda\geq 0.1$ (right).\label{Fig6} }
\end{center}
\end{figure}

Local observables therefore suggest that the model in Eq.~(\ref{eq:muller}) at $\lambda\sim 0.1$ appears thermal rather than localized. We reach similar conclusions by considering nonlocal quantities such as entanglement entropy $S_{\rm ent}$, shown in Fig.~\ref{Fig6}. Entanglement entropy $S_{ent}$ of typical many-body eigenstates is known to scale extensively with system size $L$ in the ergodic case, and in contrast obeys an area-law scaling in the MBL phase~\cite{bauer}. In Fig.~\ref{Fig6} we plot the entropy density $S_{\rm ent}/L$ for all the many-body eigenstates within a given energy window in the middle of the band at $\lambda=0.1$. Note that we shift the data for different $L$ by a certain amount $E^*$ dependent on $L$, so that the zero energy corresponds to the center of the band where states have the largest $S_{ent}$. The characteristic concave shape of entropy is suggestive of thermal behaviour; moreover, as we increase the system size, $S_{\rm ent}/L$ appears to increase, which is inconsistent with localized behaviour where $S_{\rm ent}/L$ should vanish. 

In Fig.~\ref{Fig6}(inset) we extract the average entropy $\langle S_{\rm ent} \rangle$ within the central window $|E-E^*|/L<0.02$, and perform finite-size scaling for various $\lambda$. Extremely small values of $\lambda$ such as 0.001 are consistent with area law that we expect in the MBL phase. However, for $\lambda\geq 0.1$ the average entropy within the central band clearly scales extensively with system size, suggesting that the system delocalizes.

\begin{figure*}[ttt]
  \begin{minipage}[l]{\linewidth}
\includegraphics[width=0.8\textwidth]{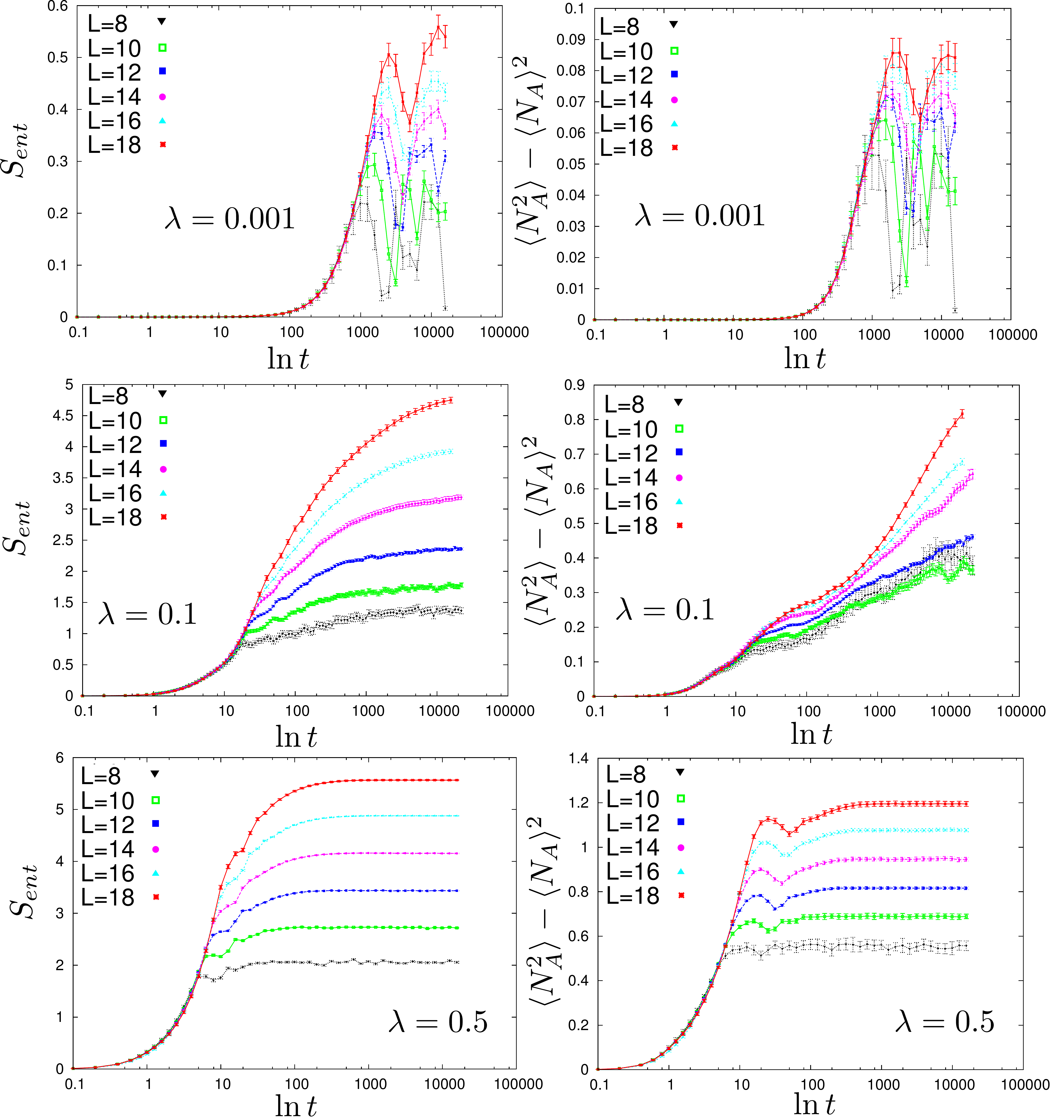}
\end{minipage}
\caption{(Color online) Growth of entanglement entropy and particle number fluctuations in the left half of the system following the quench from a product state in the model (\ref{eq:muller}). Data is averaged over random initial product states. Left panels show the entanglement entropy, right panels correspond to particle number fluctuations. Top to bottom: $\lambda=0.001$, $\lambda=0.1$ and $\lambda=0.5$. }
\label{FigA2}
\end{figure*}
Finally in Fig.~\ref{FigA2} we consider the dynamics of entanglement entropy (left panels) and the particle number fluctuations (right panels) in the left half of the system for $\lambda=0.001$, $\lambda=0.1$ and $\lambda=0.5$. Particle number fluctuations are defined as $\langle N_A^2\rangle - \langle N_A\rangle^2$, where ``A" stands for the left half of the system. It is instructive to consider both entropy and particle fluctuations because in the MBL phase with quenched disorder, the particle number fluctuations are strongly suppressed and depend weakly on system size due to the absence of transport in the system. On the other hand, the entropy still displays an unbounded logarithmic growth due to the dephasing dynamics in the MBL phase. In Fig.~\ref{FigA2} finite systems of various sizes are prepared in an initial product state and evolved unitarily in time with the Hamiltonian (\ref{eq:muller}). The data is averaged over random initial product states. Top row shows the case $\lambda=0.001$; for such a small value of $\lambda$, the systems appears localized, although upon closer inspection we see that the long-time values of entropy and particle number fluctuations still grow slowly with $L$, indicating much stronger finite size effects than in the models with quenched disorder~\cite{Serbyn13-1}. 

\begin{figure}[htb]
\includegraphics[width=\linewidth]{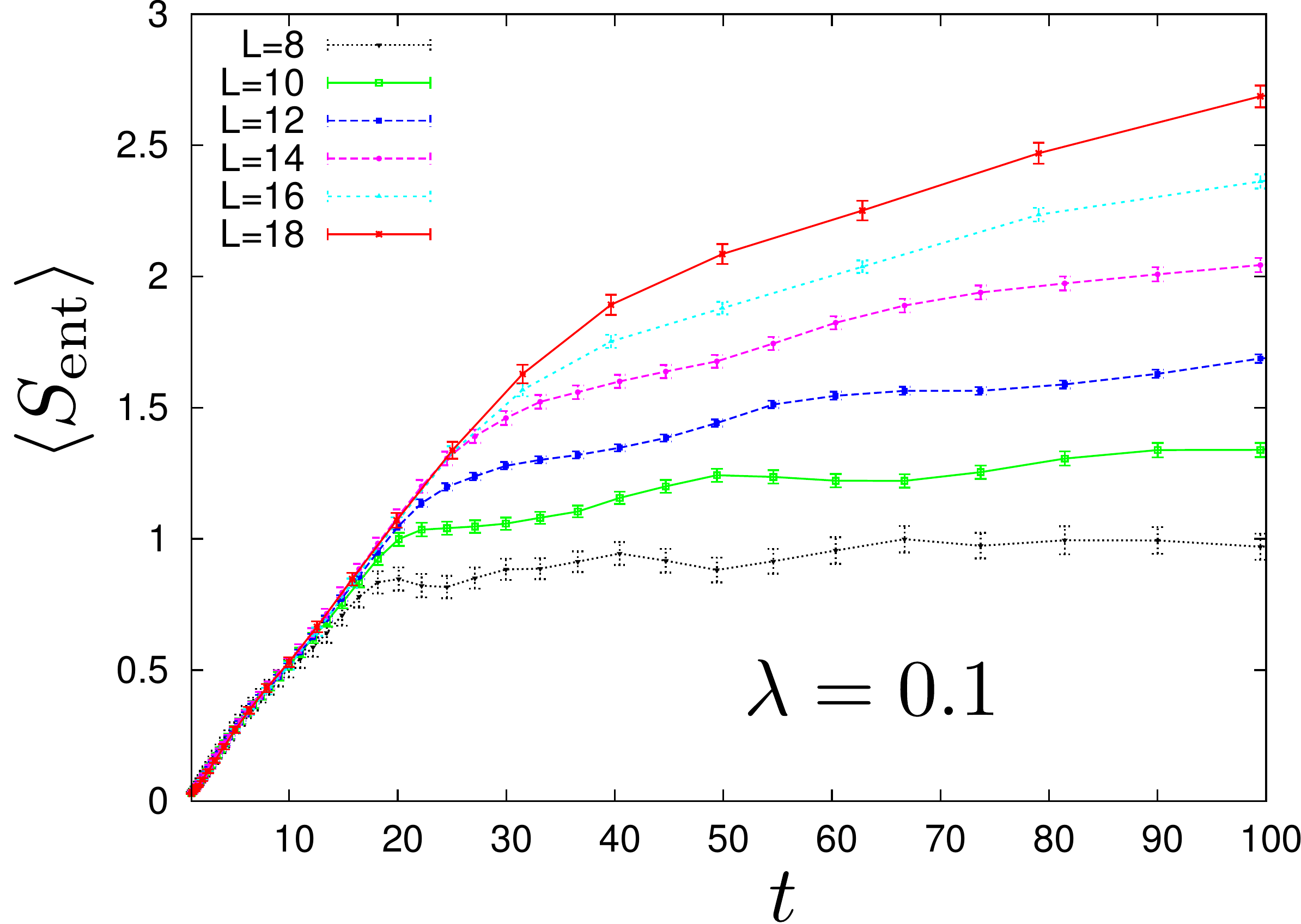}
\caption{(Color online) Growth of entanglement entropy for $\lambda=0.1$ [middle row of Fig.~\ref{FigA2}] plotted on a linear scale.}
\label{FigA3}
\end{figure}
A more interesting case is $\lambda=0.1$ shown in the middle row. If we look at $S_{\rm ent}$ for a small system such as $L=8$, we find some signatures of slow dynamics and ``$\ln t$" growth of entropy prior to saturation, as expected of the MBL behaviour~\cite{Moore12, Serbyn13-1}. However, we also notice that this slow growth of entropy is gradually washed away as we increase the system size to $L=18$. For this large system, a much faster growth of entropy sets in, until the system starts to approach the saturation value of entropy. This faster growth is linear in time, as shown in Fig.~\ref{FigA3}, and is consistent with thermalization as we take $L$ to infinity~\cite{Calabrese05, Chiara05, Kim13}. Similar conclusion can be drawn from particle number fluctuations [Fig.~\ref{FigA2}, right] which increase at long times, while they should be strongly suppressed in MBL phase with quenched disorder~\cite{Moore12}. From this, we conclude that $\lambda=0.1$ belongs to the transition regime between localized and thermal phases, where small system sizes $L$ bias the systems towards the localized regime, but for sufficiently large $L$ the system ultimately thermalizes. For comparison, we also show the case $\lambda=0.5$ [Fig.~\ref{FigA2}, bottom row] which is expected to be in the thermal phase.  

Finally one may expect in models such as (\ref{eq:muller}) that the low density of particles may favor the localization. To address the dependence of our conclusions on the filling factor, we have also analyzed the case of 1/3 filling where we found qualitatively the same phenomenology as in the half-filling case: the entropy displayed faster than logarithmic growth in time upon the increase of system size, and the particle number fluctuations also continued to grow at long times as $L$ approached the thermodynamic limit.  

\section{Conclusions} \label{sec:conc}

We studied several candidate 1D models proposed to exhibit many-body localization in the absence of quenched disorder. We emphasized that two or more very different energy scales in these models lead to pronounced finite-size effects, complicating the extrapolation of numerical results on small systems to the thermodynamic limit. More specifically, in a certain parameter regime ($\lambda\ll 1$), the DOS of small finite systems consists of bands separated by large energy gaps -- a feature that generally does not persist in the thermodynamic limit. This artificial separation reduces the ability of the system to act as a heat bath for its parts. In this regime, we found, in agreement with previous works~\cite{yao,muller2}, that the models exhibit phenomenology which is consistent with MBL. However, due to severe finite-size effects, it remains unclear whether such this type of MBL indeed persists in the thermodynamic limit.

Further, we considered the parameter range $\lambda\sim 0.1$ where finite-size effects are weaker, and numerical simulations are expected to provide more reliable information about the thermodynamic limit. In this regime, using a variety of probes, we found thermal rather than MBL behaviour. Our results show that MBL in translation-invariant systems, if it exists, is much less robust than the perturbative arguments~\cite{muller1,muller2} suggest (e.g., for the models studied here, perturbative arguments predict MBL, rather than thermal phase, at $\lambda\sim 0.1$). Further work is needed to establish the mechanism of delocalization in these models, which might have non-perturbative origin~\cite{deroeck2}. 

\section{acknowledgments}

We thank F. Huveneers, M. Lukin, J. Moore, I. Cirac, D. Huse and M. Fisher for enlightening discussions. We acknowledge support by Alfred Sloan Foundation and Early Researcher Award by the Government of Ontario (DA). Z.P. acknowledges support by DOE grant DE-SC0002140. Research at Perimeter Institute is supported by the Government of Canada through Industry Canada and by the Province of Ontario through the Ministry of Economic Development \& Innovation. This work made use of the facilities of N8 HPC provided and funded by the N8 consortium and EPSRC (Grant No.EP/K000225/1).

\end{document}